\documentclass{nature}

\usepackage{graphicx}

\usepackage{cite}

\usepackage{amsmath}
\usepackage{amssymb}

\title{Spawning rings of exceptional points out of Dirac cones}

\author{Bo Zhen$^{1\ast}$, Chia Wei Hsu$^{1,2\ast}$, Yuichi Igarashi$^{1,3\ast}$, Ling Lu$^{1}$, Ido Kaminer$^{1}$, Adi Pick$^{1,4}$, Song-Liang Chua$^{5}$,  John D. Joannopoulos$^{1}$ \& Marin Solja\v{c}i\'{c}$^{1}$}

\begin{document}

\maketitle

\vspace{24pt}

\begin{spacing}{1.5}
\noindent \normalsize{
$^{1}$Research Laboratory of Electronics, Massachusetts Institute of Technology, Cambridge, Massachusetts 02139, USA. \\
$^{2}$Department of Applied Physics, Yale University, New Haven, CT 06520, USA. \\
$^{3}$Smart Energy Research Laboratories, NEC Corporation, 34 Miyuiga-ka, Tsukuba, Ibaraki 305-8501, Japan. \\
$^{4}$Department of Physics, Harvard University, Cambridge, Massachusetts 02138, USA. \\
$^{5}$DSO National Laboratories, 20 Science Park Drive, Singapore, 118230. \\
$^\ast$These authors contributed equally to this work.}
\end{spacing}

\date{\today}

\clearpage

\begin{abstract}

The Dirac cone underlies many unique electronic properties of graphene\cite{2009_Geim_RMP} and topological insulators\cite{2010_Hasan_RMP},
and its band structure---two conical bands touching at a single point---has also been realized for photons in waveguide arrays\cite{2013_Rechtsman_NatPhoton}, atoms in optical lattices\cite{2012_Tarruell_Nature},
and through accidental degeneracy
\cite{2011_Huang_NatMat, 2013_Moitra_NatPhoton}.
Deformations of the Dirac cone often reveal intriguing properties; an example is the quantum Hall effect, where a constant magnetic field breaks the Dirac cone into isolated Landau levels\cite{2005_Zhang_Nature}.
A seemingly unrelated phenomenon is the exceptional point\cite{Moiseyev_book,Kato_book,2009_Rotter_JPA,2012_Heiss_JPA}, also known as the parity-time symmetry breaking point\cite{1998_Bender_PRL, 2008_Klaiman_PRL, 2010_Ruter_Nat.Phys,2011_Chong_PRL}, where two resonances coincide in both their positions and widths. 
Exceptional points lead to counter-intuitive phenomena such as loss-induced transparency\cite{2009_Guo_PRL}, unidirectional transmission or reflection\cite{2008_Makris_PRL,2009_Longhi_PRL,2011_Lin_PRL,2012_Regensburger_Nature,2013_Feng_NatMatt, 2014_Peng_Nat.Phys,2014_Chang_Nat.Phys}, and lasers with reversed pump dependence\cite{2012_Liertzer_PRL,2014_brandstetter_NatComm,2014_Peng_Science} or single-mode operation\cite{2014_Hodaei_Science,2014_Feng_Science}.
These two fields of research are in fact connected: here we discover the ability of a Dirac cone to evolve into a ring of exceptional points, which we call an ``exceptional ring.''
We experimentally demonstrate this concept in a photonic crystal slab.
Angle-resolved reflection measurements of the photonic crystal slab reveal that the peaks of reflectivity follow the conical band structure of a Dirac cone from accidental degeneracy, whereas the complex eigenvalues of the system are deformed into a two-dimensional flat band enclosed by an exceptional ring.
This deformation arises from the dissimilar radiation rates of dipole and quadrupole resonances, which play a role analogous  to the loss and gain in parity-time symmetric systems.
Our results indicate that the radiation that exists in any open system can fundamentally alter its physical properties in ways previously expected only in the presence of material loss and gain.

\end{abstract}

\maketitle

Closed and lossless physical systems are described by Hermitian operators, which guarantee realness of the eigenvalues and a complete set of eigenfunctions that are orthogonal to each other. 
On the other hand, systems with open boundaries\cite{2009_Rotter_JPA, 2015_Cao_RMP} or with material loss and gain\cite{1998_Bender_PRL,2009_Guo_PRL,2011_Lin_PRL,2010_Ruter_Nat.Phys,2012_Regensburger_Nature,2014_Chang_Nat.Phys,2014_Peng_Nat.Phys,2013_Feng_NatMatt,2012_Liertzer_PRL,2014_brandstetter_NatComm,2014_Peng_Science,2014_Hodaei_Science,2014_Feng_Science,2008_Klaiman_PRL,2008_Makris_PRL,2009_Longhi_PRL} are non-Hermitian\cite{Moiseyev_book} and have non-orthogonal eigenfunctions with complex eigenvalues where the imaginary part corresponds to decay or growth.
The most drastic difference between Hermitian and non-Hermitian systems is that the latter exhibit exceptional points (EPs) where both the real and the imaginary parts of the eigenvalues coalesce.
At an EP, two (or more) eigenfunctions collapse into one so the eigenspace no longer forms a complete basis, and this eigenfunction becomes orthogonal to itself under the unconjugated inner product\cite{Moiseyev_book,Kato_book,2009_Rotter_JPA,2012_Heiss_JPA}.
To date, most studies of EP and its intriguing consequences concern parity-time symmetric systems that rely on material loss and gain\cite{1998_Bender_PRL,2009_Guo_PRL,2011_Lin_PRL,2010_Ruter_Nat.Phys,2012_Regensburger_Nature,2014_Chang_Nat.Phys,2014_Peng_Nat.Phys,2013_Feng_NatMatt,2012_Liertzer_PRL,2014_brandstetter_NatComm,2014_Peng_Science,2014_Hodaei_Science,2014_Feng_Science,2008_Klaiman_PRL,2008_Makris_PRL,2009_Longhi_PRL}, but EP is a general property that requires  only non-Hermiticity.
Here, we show the existence of EPs in a photonic crystal slab with negligible absorption loss and no artificial gain.
When a Dirac-cone system has dissimilar radiation rates, the band structure is altered abruptly to show branching features with a ring of EPs.
We provide a complete picture from analytic model and numerical simulation to experimental observation; together, they illustrate the role of radiation-induced non-Hermiticity that bridges the study of EPs and the study of Dirac cones.

We start by showing that non-Hermiticity from radiation can deform an accidental Dirac point into a ring of EPs.
First, consider a 2D photonic crystal (PhC)\cite{JJ_book} (inset of Fig.~1a), where a square lattice (periodicity $a$) of circular air holes (radius $r$) is introduced in a dielectric material. This is a Hermitian system, as there is no material gain or loss and no open boundary for radiation.
By tuning a system parameter (for example, $r$), one can achieve accidental degeneracy between a quadrupole mode and two degenerate dipole modes at the $\Gamma$ point (center of the Brillouin zone), leading to a linear Dirac dispersion due to the anti-crossing between two bands with the same symmetry\cite{2011_Huang_NatMat, 2012_Sakoda_OE}. 
The accidental Dirac dispersion from the effective Hamiltonian model (see equation~\eqref{eqn:NH-Hamiltonian} below with $\gamma_{0}=0$) is shown as solid lines in Fig.~1a , agreeing with numerical simulation results (symbols in Fig.~1a).
In the effective Hamiltonian we do not consider the dispersionless third band (gray line) due to symmetry arguments (section I in Supplementary Information), although this third band cannot be neglected in certain calculations, including Berry phase and effective medium property\cite{2012_Chan_Review, 2012_Mei_PRB}. 

Next, we consider a similar, but open, system: a PhC slab (inset of Fig.~1b) with finite thickness $h$. With the open boundary, modes within the radiation continuum become resonances because they radiate by coupling to extended plane waves in the surrounding medium.
Non-Hermitian perturbations need to be included in the Hamiltonian to account for the radiation loss.
To the leading order, radiation of the dipole mode can be described by adding an imaginary part $-i\gamma_{\rm d}$ to the Hamiltonian, while the quadrupole mode does not radiate due to its symmetry mismatch with the plane waves\cite{2012_Lee_PRL}.
Specifically, at the $\Gamma$ point the system has $C_2$ rotational symmetry (invariant under $180^\circ$  rotation around the $z$ axis), and the quadrupole mode does not couple to the radiating plane wave because the former is even [${\bf E}({\bf r}) = \hat{O}_{C_{2}} {\bf E}({\bf r}) $] whereas the latter is odd  [${\bf E}({\bf r}) = -\hat{O}_{C_{2}} {\bf E}({\bf r}) $] under $C_2$ rotation\cite{JJ_book}.
The effective Hamiltonian is
\begin{equation} 
\label{eqn:NH-Hamiltonian}
H_{\rm eff} = \left( \begin{array}{cc}
\omega_{0} & v_{g}k  \\
v_{g}k & \omega_{0}-i\gamma_{\rm d} \end{array} \right),
\end{equation} 
with complex eigenvalues
\begin{equation} 
\label{eqn:EPring}
\omega_{\pm} = \omega_{0} - i\frac{\gamma_{\rm d}}{2} \pm v_{g}\sqrt{k^{2}-k_{c}^{2}},
\end{equation} 
where $\omega_{0}$ is the frequency at accidental degeneracy, $v_{g}$ is the group velocity of the linear Dirac dispersion in the absence of radiation, $k$ is the magnitude of the in-plane wavevector ($k_{x},k_{y}$), 
and $k_{c} \equiv \gamma_{\rm d}/2v_{g}$.
Here, one of the three bands is decoupled from the other two and is not included in equation~\eqref{eqn:NH-Hamiltonian} (see section II of Supplementary Information).
In equation~\eqref{eqn:EPring}, 
a ring defined by $k=k_{c}$ separates the $k$ space into two regions:
inside the ring ($k<k_{c}$), Re($\omega_{\pm}$) are dispersionless and degenerate; outside the ring ($k>k_{c}$), Im($\omega_{\pm}$) are dispersionless and degenerate.
In the vicinity of $k_{c}$, Im($\omega_{\pm}$) and Re($\omega_{\pm}$) exhibit square-root dispersion (also known as branching behavior) inside and outside the ring, respectively. 
Exactly on the ring ($k=k_{c}$), the two eigenvalues $\omega_{\pm}$ are degenerate in both real and imaginary parts; meanwhile, the matrix $H_{\rm eff}$ becomes defective with an incomplete eigenspace spanned by only one eigenvector (1, -$i$)$^{T}$ that is orthogonal to itself under the unconjugated inner product.
This self-orthogonality is the definition of EPs; hence, here we have not just one EP, but a continuous ring of EPs. We call it an {\it exceptional ring}.

Fig.~1b,c show the complex eigenvalues of the PhC slab structure calculated numerically (symbols), which closely follow the analytic model of equation~\eqref{eqn:EPring} shown as solid lines in the figure.
When the radius $r$ of the holes is tuned away from accidental degeneracy, the exceptional ring and the associated branching behavior disappear, as shown in Fig.~S1. Several properties of the PhC slab contribute to the existence of this exceptional ring.
Due to periodicity, one can probe the dispersion from two degrees of freedom, $k_x$ and $k_y$, in just one structure.
The open boundary provides radiation loss, and the $C_2$ rotational symmetry differentiates the radiation loss of the dipole mode and of the quadrupole mode.

We can rigorously show that the exceptional ring exists in realistic PhC slabs, not just in the effective Hamiltonian model. 
Our proof is based on the unique topological property of EPs: 
when the system parameters evolve adiabatically along a loop encircling an EP, the two eigenvalues switch their positions when the system returns to its initial parameters\cite{2009_Rotter_JPA, 2001_Dembowski_PRL, 2012_Heiss_JPA,2015_Cao_RMP}, in contrast to the typical case where the two eigenvalues return to themselves. 
Using this property, we numerically show, in Fig.~S2  and section III of Supplementary Information, that the complex eigenvalues always switch their positions along every direction in the $k$ space, and therefore prove the existence of this exceptional ring.
As opposed to the simplified effective Hamiltonian model, in a real PhC slab, the EP may exist at a slightly different magnitude of $k$ and for a slightly different hole radius $r$ along different directions in the momentum space, but this variation is small and negligible in practice (section IV of Supplementary Information).

To demonstrate the existence of the exceptional ring in such a system, we fabricate large-area periodic patterns in a Si$_{3}$N$_{4}$ slab ($n=2.02$, thickness 180 nm) on top of 6 $\mu$m of silica ($n=1.46$) using interference photolithography\cite{2012_Lee_PRL}. 
Scanning electron microscope (SEM) images of the sample are shown in Fig.~2a, featuring a square lattice (periodicity $a=336$ nm) of air cylindrical holes with radius of $109$ nm. 
We immerse the structure into an optical liquid and tune the refractive index of the liquid; accidental degeneracy in the Hermitian part is achieved when the liquid index is selected to be $n= 1.48$.
We perform angle-resolved reflectivity measurements (setup shown in Fig.~2b) between 0 and 2 degrees along the $\Gamma$ to X direction and the $\Gamma$ to M direction, for both $s$ and $p$ polarizations. 
The measured reflectivity for the relevant polarization is  plotted in the upper panel of Fig.~2c, showing good agreement with numerical simulation results (lower panel),
with differences coming from scattering of disorder, inhomogeneous broadening, and 
the uncertainty in the measurements of system parameters.
The complete experimental result for both polarizations is shown in Fig.~S3; the third and dispersionless band shows up in the other polarization, decoupled from the two bands of interest.

The peaks of reflectivity (dark red color in Fig.~2c) follow the linear Dirac dispersion; this feature disappears for structures with different radii that do not reach accidental degeneracy (experimental results in Fig.~S4).
To understand the reflection peaks, we consider a generic two-by-two Hamiltonian $H$ with no assumption made about its matrix elements.
We separate $H$ into a Hermitian part A and an anti-Hermitian part -$i$B (so that A and B are both Hermitian), and choose the basis in which A is diagonal:
\begin{equation}
\label{eq:TCMT}
UHU^{\rm T} = \underbrace{\left( \begin{array}{cc}
\Omega_{1} & 0 \\
0 & \Omega_{2} \end{array} \right)
}_{\hbox{A}} 
-
\underbrace{ i
\left( \begin{array}{cc}
\gamma_{1} & \gamma_{12}  \\
\gamma_{12}^* & \gamma_{2} \end{array} \right) 
}_{\hbox{{\it i}B}}
\xrightarrow{\text{eigenvalues}}
\left( \begin{array}{cc}
\omega_{+} & 0  \\
0 & \omega_{-}  \end{array} \right).
\end{equation}
As before, we use $\omega_{\pm}$ to denote the complex eigenvalues of the Hamiltonian $\hbox{A}-i\hbox{B}$.
The reflectivity in our system can be modeled using temporal coupled-mode theory (TCMT, with details in section V of Supplementary Information), where we show that the reflection peaks generally occur near the eigenvalues $\Omega_{1,2}$ of the Hermitian part A and are independent of the non-Hermitian part $-i$B (Fig.~S5 with details in section VI in Supplementary Information).
Therefore, the peak locations in Fig.~2c (dark red) reveal information about only the Hermitian part of the Hamiltonian; the fact that they show linear Dirac dispersion indicates that we have successfully achieved accidental degeneracy in the eigenvalues of the Hermitian part, consistent with the simplified model in equation~\eqref{eqn:NH-Hamiltonian}.
In Fig.~S6, we plot the $\Omega_{1,2}$ extracted from the reflectivity data through a more rigorous data analysis using TCMT (described below); the linear dispersion is indeed observed.

The eigenvalues of the Hamiltonian, $\omega_{\pm}$, behave very differently from the reflectivity peaks. 
Simulation results  (white lines in the lower panel of Fig.~2c) show Re$(\omega_{\pm})$ are dispersionless at small angles with a branch-point singularity around $0.31^{\circ}$---consistent with the feature predicted by the simplified Hamiltonian in equation~\ref{eqn:EPring}.
In Fig.~2d, we compare the reflectivity spectra from simulations (with peaks indicated in red arrows)  with the corresponding complex eigenvalues at three representative angles ($0.8^{\circ}$ in blue, $0.31^{\circ}$ in green, and $0.1^{\circ}$ in magenta).
At $0.31^{\circ}$, the two complex eigenvalues are degenerate, indicating an EP; however, the two reflection peaks do not coincide since they  represent the eigenvalues of only the Hermitian part of the Hamiltonian, which does not have degeneracy here.
The dip in reflectivity between the two peaks (marked as black arrows in Figs.~2 and 3) is the coupled-resonator-induced transparency (CRIT) that arises from the interference between radiation of the two resonances\cite{2010_Miroshnichenko_RMP, 2014_Hsu_NL}, similar to electromagnetically induced transparency (EIT)\cite{2005_Fleischhauer_RMP}.

To extract the underlying Hamiltonian matrix and its eigenvalues from the measured reflectivity spectrum, we use TCMT to model the direct and the resonant reflection processes; the expression for reflectivity is given in equation~\eqref{eq:R} with the full derivation given in section V of the Supplementary Information.
Fitting the reflectivity curves with the TCMT expression gives us the matrix elements of the Hamiltonian (as shown in equation~\eqref{eq:TCMT}) that we use to calculate its eigenvalues; this procedure is the same as our approach in Ref.~\citen{2013_Hsu_Nature} except that here we handle multiple resonances simultaneously, accounting for their non-orthogonality and radiative coupling\cite{2004_Suh_IEEE}.
Fig.~3a compares the fitted and the measured reflectivity curves at three representative angles (with more comparison in Fig.~S6a); the excellent agreement shows the validity of the TCMT equations. Underneath the reflectivity curves, we show the complex eigenvalues.

Repeating the fitting procedure for reflectivity spectrum measured  at different angles, we obtain the dispersion curves for all complex eigenvalues, which are plotted in Fig.~3b.
Along both directions in $k$ space ($\Gamma \rightarrow$ X and $\Gamma \rightarrow$ M), the two bands of interest (shown in blue and red) exhibit the EP behavior predicted in equation~\eqref{eqn:EPring}: 
for $k<k_{c}$ the real parts are degenerate and dispersionless;
for $k>k_{c}$ the imaginary parts are degenerate and dispersionless;
for $k$ in the vincinity of $k_{c}$ branching features are observed in the real or imaginary part. 
In Fig.~3c, we plot the eigenvalues on the complex plane for both the $\Gamma \rightarrow$ X and $\Gamma \rightarrow$ M directions. We can see that in both directions, the two eigenvalues approach each other and become very close at certain $k$ point, which is a clear signature of the system being very near EP.

We have shown that non-Hermiticity arising from radiation can significantly alter fundamental properties of the system including the band structures and density of states; this effect becomes most prominent near EPs.  
The PhC slab described here provides a simple-to-realize platform for studying the influence of EPs on light-matter interaction,
such as for single particle detection\cite{2014_Wiersig_PRL} and modulation of quantum noise\cite{2011_Yoo_PRA}.
The two-dimensional flat band also provides high density of states and therefore high Purcell factors. 
The strong dispersion of loss in the vicinity of the $\Gamma$ point can improve the performance of large-area single-mode PhC lasers\cite{2014_Song_OL}.
The deformation into exceptional ring can also occur for non-accidental Dirac points\cite{2011_Szameit_PRA}.
Further studies can advance the understanding of the connection between the topological property of Dirac points\cite{2010_Hasan_RMP, 2014_Lu_nphoton} and that of EPs\cite{2001_Dembowski_PRL} in general non-Hermitian wave systems, and this method of our study goes beyond photonics to phonons, electrons, and atoms. 

\noindent{\bf \large METHODS SUMMARY}

\noindent {\bf Sample fabrication.}
The Si$_3$N$_4$ layer was grown with the low-pressure chemical vapor deposition method on a 6$\mu$m-thick cladding of SiO$_2$ on the backbone of a silicon wafer (LioniX). Before exposure, the wafer was coated with a layer of polymer as anti-reflection coating, a thin layer of SiO$_2$ as an intermediate layer for etching, and a layer of negative photoresist for exposure. The square lattice pattern was created with Mach$-$Zehnder interference lithography using a 325-nm He/Cd laser. The angle between the two arms of the laser beam was chosen for a periodicity of 336 nm. After exposures, the pattern in the photoresist was transferred to Si$_3$N$_4$ by reactive-ion etching. 

\noindent {\bf Experimental details.}
The source was a supercontinuum laser from NKT Photonics (SuperKCompact). A polarizer selected $s$- or $p$-polarized light. The sample was immersed in a colorless liquid with tunable refractive indices (Cargille Labs). The sample was mounted on two perpendicular motorized rotation stages (Newport): one to orient the PhC to the $\Gamma$-X or $\Gamma$-M direction, and the other to determine the incident angle $\theta$. The reflectivity spectra were measured with a spectrometer with spectral resolution of 0.02 nm (HR4000; Ocean Optics). 

\centerline{\rule{0.8 \columnwidth}{1pt}}

\begin{addendum}
 \item[Supplementary Information] is available in the online version of the paper.
 \item[Acknowledgments] 
The authors thank Dr. Tim Savas for fabrication of the samples. Also, the authors thank Fan Wang, Yi Yang, Nick Rivera, Scott Skirlo, Dr. Owen Miller, 
and Prof. Steven G. Johnson for helpful discussions. This work was partly supported by the Army Research Office through the Institute for Soldier Nanotechnologies under contract no. W911NF-07-D0004 and no. W911NF-13-D-0001. B.Z., L.L., and
M.S. were partly supported by S3TEC, an Energy Frontier Research Center funded by
the US Department of Energy under grant no. DE-SC0001299. L.L. was supported in part by the Materials Research Science and Engineering Center of the National Science Foundation (award no. DMR-1419807). I.K. was supported in part
by Marie Curie grant no. 328853-MC-BSiCS.
 \item[Author Contributions] 
All authors discussed the results and made critical contributions to the work.
 \item[Author Information] Reprints and permissions information is available at www.nature.com/reprints. The authors declare no competing financial interests. Correspondence and requests for materials should be addressed to B.Z.~(email: bozhen@mit.edu).
 \item[Competing financial interests]
The authors declare no competing financial interests.
\end{addendum}

\clearpage

\begin{figure*}[ht]
\includegraphics[width=\textwidth]{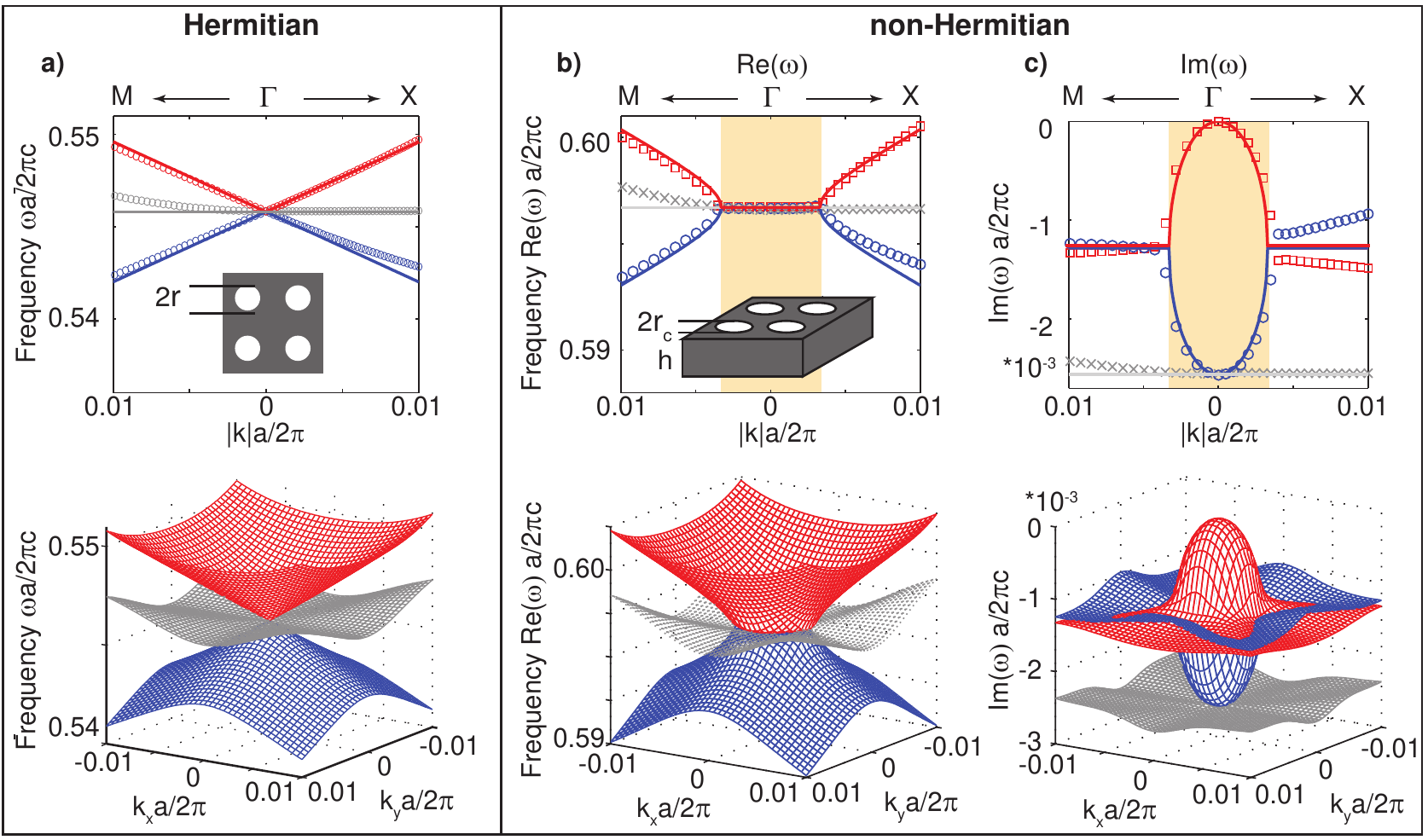}
\caption{\baselineskip20pt{\bf Accidental degeneracy in Hermitian and non-Hermitian photonic crystals (PhC).}
{\bf a}, Band structure of a 2D PhC consisting of a square lattice of circular air holes. Tuning the radius $r$ leads to accidental degeneracy between a non-degenerate quadrapole band and two doubly degenerate dipole bands, resulting in two bands with linear Dirac dispersion (red and blue) and a flat band (gray).
{\bf b},{\bf c}, The real and imaginary parts of the eigenvalues of an open, and therefore non-Hermitian, system: a PhC slab with finite thickness $h$. By tuning the radius, accidental degeneracy in the real part can be achieved, but the Dirac dispersion is deformed due to the non-Hermiticity. 
The analytic model predicts that the real (imaginary) part of the eigenvalue stays as a constant within (outside) a ring in the wavevector space, indicating two flat bands in dispersion, with a ring of exceptional points (EPs) where both the real and the imaginary parts are degenerate.
In the upper panels, solid lines are from the analytic model and symbols are from numerical simulations: red squares represent the band connecting to the quadrapole mode at the center; blue circles represent the band connecting to the dipole mode at the center; and gray crosses represent the third band that is decoupled from the previous two due to symmetry. The 3D plots in the lower panels are from simulations.
}
\label{fig1}
\end{figure*}

\begin{figure*}[ht]
\centerline{
\includegraphics[width=\textwidth]{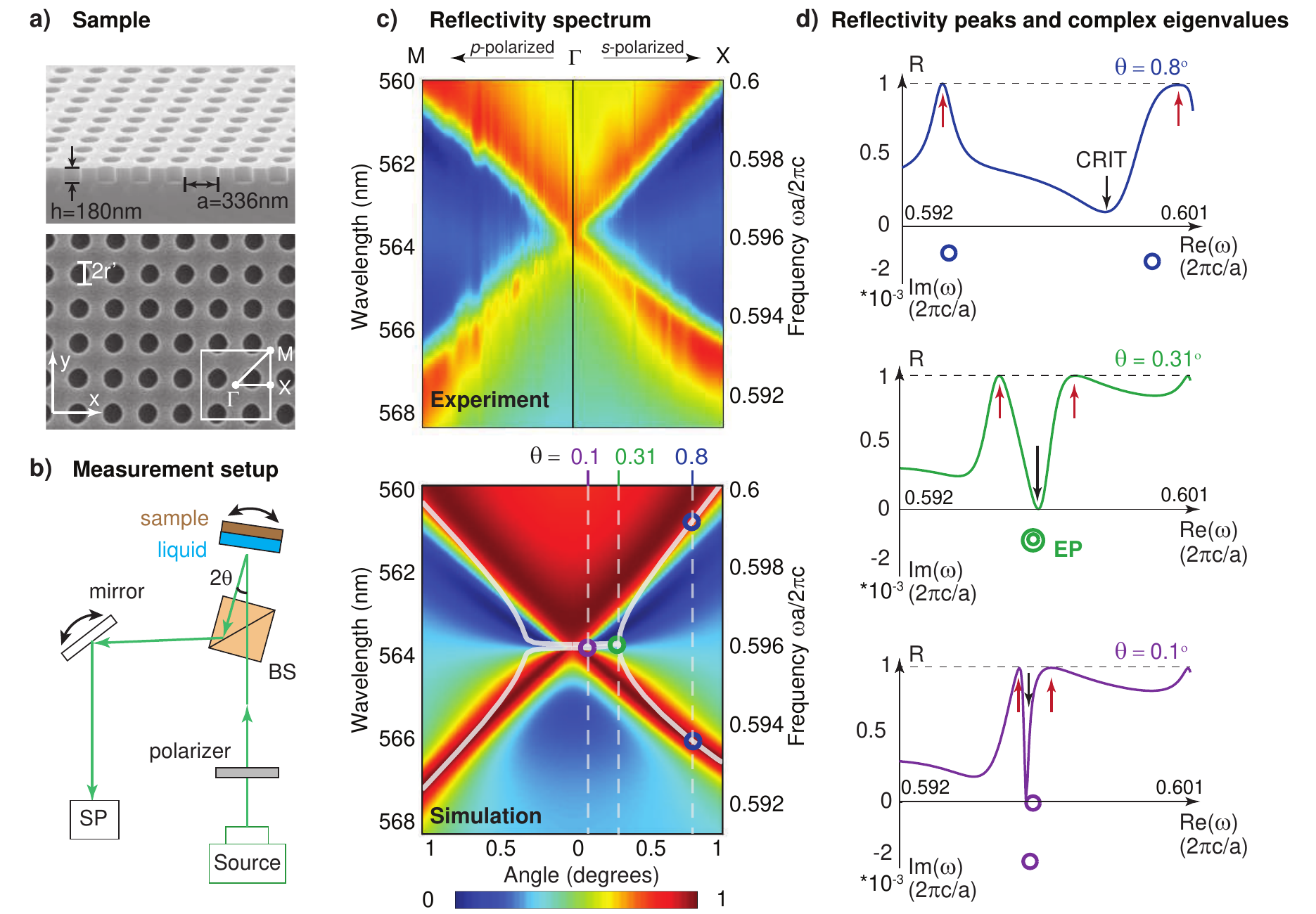}
}\caption{\baselineskip19pt{\bf Experimental reflectivity spectrum and accidental Dirac dispersion.}
{\bf a}, SEM images of the PhC samples: side view (upper panel) and top view (lower panel). 
{\bf b}, Schematic drawing of the measurement setup. Light from a super-continuum source reflects off the PhC slab and is collected using a spectrometer. The incident angle is controlled using a precision rotationary stage. (BS: beam splitter; SP: spectrometer)
{\bf c}, Reflectivity spectrum of the sample measured experimentally (upper panel) and calculated numerically (lower panel) 
along the $\Gamma$ to X and the $\Gamma$ to M directions.
The peak location of reflectivity reveals the Hermitian part of the system, which forms Dirac dispersion due to accidental degeneracy.
White lines in the lower panel indicate real part of the eigenvalues. 
{\bf d}, Three line cuts of reflectivity from simulation results.
Also shown are the complex eigenvalues (hollow circles) calculated numerically. At large angles (0.8$^{\circ}$), the two resonances are far apart, so the reflectivity peaks (red arrows) are close to the actual positions of the complex eigenvalues. However, at small angles (0.3$^{\circ}$, 0.1$^{\circ}$), the coupling between resonances cause the resonance peaks (red arrows) to have much greater separations in frequencies compared to the complex eigenvalues. The black arrows mark the dips in reflectivity that correspond to the coupled-resonator induced transparency (CRIT, see text for details). 
}
\label{fig2}
\end{figure*}

\setcounter{figure}{2}
\begin{figure*}[ht]
\begin{centering}
\includegraphics[width=\textwidth]{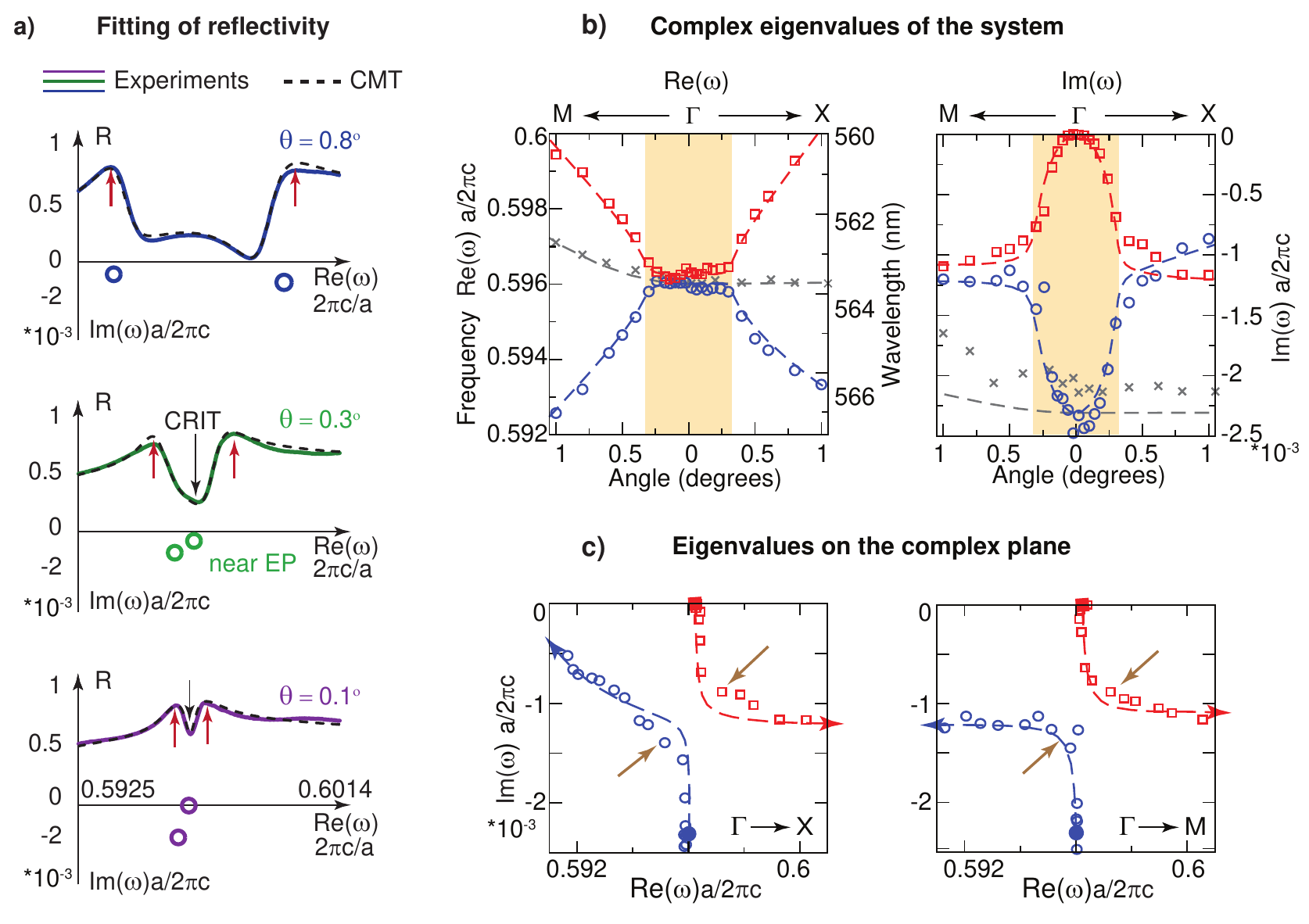}
\caption{\baselineskip20pt{\bf Experimental demonstration of exceptional ring.}
{\bf a}, Examples of reflection spectrum from the sample at three different angles (0.8$^{\circ}$ blue, 0.3$^{\circ}$ green and 0.1$^{\circ}$ magenta, solid lines) measured with $s$-polarized light along $\Gamma$ to X direction (same setup as in numerical simulations shown in Fig.~2d), fitted with the TCMT expression (equation~\eqref{eq:R}) (black dashed lines). At each angle, the position of the complex eigenvalue extracted experimentally are shown as hollow circles.
{\bf b}, Complex eigenvalues extracted experimentally (symbols), with comparison to numerical simulation results (dashed lines) for both the real part (left panel) and the imaginary part (right panel).
Red squares and dashed lines are used for the band with zero radiation loss at the $\Gamma$ point, blue circles and dashed lines for the band with finite radiation loss at the $\Gamma$ point, and gray crosses and dashed lines for the third band decoupled from the previous two due to symmetry.
{\bf c}, Positions of the eigenvalues approach and become very close to each other (indicated by the two brown arrows), demonstrating near EP features in different directions in the momentum space and the existence of an exceptional ring.    
}
\label{fig3}
\end{centering}
\end{figure*}
\clearpage

\noindent {\bf \Large Supplementary information}


\renewcommand{\theequation}{S.\arabic{equation}}
\setcounter{equation}{0}


\subsection{Section I. Effective Hamiltonian of accidental Dirac points in Hermitian systems\\}	
To the leading order of approximation, the effective Hamiltonian for accidental Dirac cones in Hermitian systems (2D PhC) is written as a $3\times3$ matrix due to the involvement of three bands: 
\begin{equation}
\label{eqn:H-Hamiltonian}
H_{\rm eff}^{\rm 2D} =
\left( \begin{array}{ccc}
\omega_{0} & v_{g}k_{x} & v_{g}k_{y} \\
v_{g}k_{x} & \omega_{0} & 0 \\
v_{g}k_{y} & 0 & \omega_{0} \end{array} \right) 
\end{equation}
that can be transformed into: 
\begin{equation}
\label{eqn:transform-Dirac}
U H_{\rm eff}^{\rm 2D} U^{\rm T} =
\left( \begin{array}{ccc}
\omega_{0} & v_{g}k & 0 \\
v_{g}k & \omega_{0} & 0 \\
0 & 0 & \omega_{0} \end{array} \right) 
\end{equation}
with the orthogonal transformation matrix 
\begin{equation} 
\label{eqn:unitary}
U= \left( \begin{array}{ccc}
1 & 0 & 0 \\
0 & \cos{\theta} & \sin{\theta} \\
0 & -\sin{\theta} & \cos{\theta} \end{array} \right)
\end{equation}
Here, $\cos{\theta}=k_{x}/k$, $\sin{\theta}=k_{y}/k$, $k_{x,y}$ are in-plane wavevectors.
After transformation, the $3\times3$ matrix becomes two isolated blocks: the upper $2\times 2$ block gives the conical Dirac dispersion ($\omega = \omega_{0} \pm v_{g}k$), while the lower block is the intersecting flat band ($\omega = \omega_{0}$).

\subsection{Section II. Effective non-Hermitian Hamiltonian of the exceptional ring\\}
For a 3D PhC slab that has finite thickness, the two dipole modes become resonances with finite lifetime due to their coupling to radiation; therefore, their eigenvalues become complex ($\omega_{0} - i\gamma_{\rm d}$). With $C_{4}$ rotational symmetry, these two dipole modes are identical to each other under an $90^{\circ}$ rotation and therefore share the same complex eigenvalue. Meanwhile, the quadrapole mode does not couple to radiation at the $\Gamma$ point due to symmetry mismatch, and to leading order its eigenvalue remains at $\omega_{0}$. The effective non-Hermitian Hamiltonian of the 3D PhC slab becomes
\begin{equation}
H_{\rm eff}^{\rm 3D} =
\left( \begin{array}{ccc}
\omega_{0} & v_{g}k_{x} & v_{g}k_{y} \\
v_{g}k_{x} & \omega_{0}-i\gamma_{\rm d} & 0 \\
v_{g}k_{y} & 0 & \omega_{0}-i\gamma_{\rm d} \end{array} \right),
\end{equation}
which transforms to
\begin{equation}
\label{eq:Heff3dU}
U H_{\rm eff}^{\rm 3D} U^{\rm T}  =
\left( \begin{array}{ccc}
\omega_{0} & v_{g}k & 0 \\
v_{g}k & \omega_{0}-i\gamma_{\rm d} & 0 \\
0 & 0 & \omega_{0}-i\gamma_{\rm d} \end{array} \right) 
\end{equation}
with the same matrix $U$ as in equation~\ref{eqn:unitary}.
The upper $2\times2$ block is the $H_{\rm eff}$ we refer to in equation~\eqref{eqn:NH-Hamiltonian} that givies rise to an exceptional ring, while the lower block is the intersecting flat band. 

\subsection{Section III, Existence of exceptional points along every direction in momentum space\\}

In this section, we demonstrate that EPs exist in all directions in the $k$ space, not only for a simplified Hamiltonian (equation~\ref{eqn:NH-Hamiltonian}), but also for realistic structures. 
To prove their existence, we use the unique topological property of EPs: when the system evolves adiabatically in the parameter space around an EP, the eigenvalues will switch their positions at the end of the loop\cite{2009_Rotter_JPA, 2001_Dembowski_PRL}. 
In our system, the parameter space in which we choose to evolve the eigenfunctions is three-dimensional, consisting of the two in-plane wavevectors $(k_{x},k_{y})$ and the radius of the air holes $r$, as shown in Fig.~S2a. Here, $r$ can also be other parameters, like the refractive index of the PhC slab ($n$), the periodicity of the square lattice ($a$), or the thickness of the slab ($h$). For simplicity of this demonstration, we choose $r$ as the varying parameter while keeping all other parameters ($n$, $a$, and $h$) fixed throughout. 

First, we compare the evolution of the eigenvalues when the system parameters follow (1) a loop that does not enclose an EP, and (2) one that encloses an EP.  Following the loop $A\rightarrow B\rightarrow C\rightarrow D\rightarrow A$ in Fig.~S2a,b that does not enclose an EP (point ${\rm E_{P}}$), we see that the complex eigenvalues come back to themselves at the end of the loop (Fig.~S2c where the red dot and the blue dot return to their initial positions at the end of the loop). However, following the loop $A'\rightarrow B'\rightarrow C'\rightarrow D'\rightarrow A'$ in Fig.~S2d, which encloses an EP (point ${\rm E_{P}}$), we see that the complex eigenvalues switch their positions in the complex plane (Fig.~S2f where the red dot and the blue dot switch their positions).
This switching of the eigenvalues shows the existence of an EP along the $\Gamma$ to X direction, at some particular value of $k_x$ and some particular value of radius $r$.
This shows the existence of the EP without having to locate the exact parameters of $k_x$ and $r$ at which the EP occurs. 
 
Similarly, we can evolve the parameters along any direction $\theta = {\rm tan}(k_y/k_x)$ in the $k$ space and check if an EP exists along this direction or not. 
 As two examples, we show the evolution of the complex eigenvalues when we evolve the parameters along the $\theta = \pi/8$ direction following the loop $A''\rightarrow B''\rightarrow C''\rightarrow D''\rightarrow A''$ and along the $\theta = \pi/4$ direction following the loop  $A'''\rightarrow B'''\rightarrow C'''\rightarrow D'''\rightarrow A'''$ in Fig.~S2g,h.
In both cases, we observe the switching of the eigenvalues, showing the existence of an EP along these two directions.
The same should hold for every direction in $k$ space.


The above calculations show that for every direction $\theta$ we examined in the $k$ space, there is always a particular combination of $k_c$ and $r_c$, which supports an EP. However, we note that in general, different directions can have different $k_c$ and different $r_c$, so the exceptional ring for the realistic PhC slab structure is parameterized by $k_c(\theta)$ and $r_c(\theta)$. This angular variation of $k_c(\theta)$ and $r_c(\theta)$ can be described by introducing higher order corrections in the effective Hamiltonian, which we examine in the next section.

\subsection{Section IV, Generalization of the effective Hamiltonian\\}

%
%
Here, we generalize the effective Hamiltonian in equation \eqref{eqn:NH-Hamiltonian} and  \eqref{eq:Heff3dU}.
First, the radiation of the quadrapole mode is zero only at the $\Gamma$ point; away from the $\Gamma$ point, the quadrapole mode has a  $\vec{k}$-dependent radiation that is small but non-zero, which we denote with  $\gamma_{\rm q}$.
Second, we consider possible deviation from accidental degeneracy in the Hermitian part, with a frequency walk-off $\delta$.
With these two additional ingredients, the effective Hamiltonian becomes
\begin{equation}
\label{eqn:neardegeneracy-with-loss} 
\left( \begin{array}{cc}
\omega_{0}+\delta & v_{g}k \\
 v_{g}k & \omega_{0}  \end{array} \right) \
- i
\left( \begin{array}{cc}
\gamma_{\rm q}& \sqrt{\gamma_{\rm q}\gamma_{\rm d}} \\
\sqrt{\gamma_{\rm q}\gamma_{\rm d}} &  \gamma_{\rm d}  \end{array} \right)\
,
\end{equation}
with complex eigenvalues of 
\begin{equation}
\label{eqn:neardegeneracy-with-loss-eigen} 
\omega_{\pm} = \omega_{0} +\frac{\delta}{2} - i\frac{\gamma_{\rm q}+\gamma_{\rm d}}{2}\pm \sqrt{\left(v_{g}k-i\sqrt{\gamma_{\rm q}\gamma_{\rm d}}\right)^2-\left(\frac{\gamma_{\rm d}-\gamma_{\rm q}}{2}-i\frac{\delta}{2}\right)^2},
\end{equation}
which generalizes equations~\eqref{eqn:NH-Hamiltonian} and \eqref{eqn:EPring}. 
We note that the off-diagonal term $\sqrt{\gamma_{\rm q}\gamma_{\rm d}}$ in equation~\eqref{eqn:neardegeneracy-with-loss} is required by energy conservation and time-reversal symmetry\cite{2004_Suh_IEEE, 2013_Hsu_LSA}, as we will discuss more in the next section.
Equation~\eqref{eqn:neardegeneracy-with-loss-eigen} shows that EP occurs when the two conditions
\begin{equation}
\label{eqn:EP-condition}
\begin{cases} 
k = (\gamma_{\rm d}-\gamma_{\rm q})/(2v_{g}) \approx \gamma_{\rm d}/2v_{g}, \\ 
\delta = 2 \sqrt{\gamma_{\rm d}\gamma_{\rm q}},   
\end{cases}
\end{equation}
are satisfied.
In the region of momentum space of interest, $\gamma_{\rm q}$ is much smaller than $\gamma_0$ (this can be seen, for example, from the imaginary parts of Fig.~S1a,c), so the first condition becomes $k_{\rm c} \approx \gamma_{\rm d}/2v_{g}$, same as in the simplified model.
For a given direction $\theta$ in the $k$ space (as discussed in the previous section), we can vary the magnitude $k$ and the radius $r$ to find the $k_c(\theta)$ and $r_c(\theta)$ where these two conditions are met simultaneously.


We can now analyze the angular dependence of $k_c(\theta)$ and $r_c(\theta)$ without having to find their exact values.
The first condition of equation~\eqref{eqn:EP-condition} says that the angular dependence of $k_c(\theta)$ comes from $\gamma_{\rm d}$ and $v_g$; in the PhC slab structure here, we find that $\gamma_{\rm d}$ varies by about 20\% as the angle $\theta = {\rm tan}(k_y/k_x)$ is varied; while $v_g$ remains almost the same; therefore, $k_c(\theta)$ potentially varies by around 10\% along the exceptional ring.
For the second condition of equation~\eqref{eqn:EP-condition}, we have $\gamma_{\rm d} \approx 5\times 10^{-3}\omega_{0}$ and  $\gamma_{\rm q} \approx 5\times 10^{-5}\omega_{0}$ for our PhC slab structure, so $\delta_c = 2 \sqrt{\gamma_{\rm d}\gamma_{\rm q}} \approx 1\times 10^{-3}\omega_{0}$.
Again, $\gamma_{\rm d}$ and  $\gamma_{\rm q}$ vary by about 20\% as the angle $\theta$ is changed, so $\delta_c$ can change by around $2 \times 10^{-4}\omega_{0}$.
Empirically, we find that a change of $\delta$ by $2 \times 10^{-4}\omega_{0}$ corresponds to a change in the radius $r$ of around $0.06$ nm, which is the estimated range of variation for $r_c(\theta)$ of all $\theta \in [0,2\pi)$.
This angular variation is much smaller than our structure can resolve in practice, since the radii of different holes within one fabricated PhC slab will already differ by more than $0.06$ nm.
So, in practice a given fabricated structure can be close to EP along all different directions $\theta$, but is unlikely to be an exact EP for any direction.

\subsection{Section V, Temporal Coupled Mode Theory (TCMT)\\} 

To connect the Hamiltonian of the resonances to the experimentally measured reflectivity, we resort to temporal coupled-mode theory (TCMT)\cite{Haus_book, JJ_book}. Here, we consider a very general setup with an arbitrary number of resonances in the PhC slab.
The time evolution of these $n$ resonances, whose complex amplitudes are denoted by an $n \times 1$ column vector $A$, is described by the Hamiltonian $H$ and a driving term,
\begin{equation}
\label{eq:CMT_1}
\frac{dA}{dt} = -i H A + K^{\rm T} s_+,
\end{equation}
where the Hamiltonian is an $n \times n$ non-Hermitian matrix
\begin{equation}
\label{eq:H_TCMT}
H = \Omega -i \Gamma -i \gamma_{\rm nr},
\end{equation}
with $\Omega$ denoting its Hermitian part, $-i \Gamma$ denoting its anti-Hermitian part from radiation loss, and $-i \gamma_{\rm nr}$ its anti-Hermitian part from non-radiative decays including absorption and surface roughness. For simplicity, we consider the same non-radiative loss for all resonances, so $\gamma_{\rm nr}$ is a real number instead of a matrix.

Reflectivity measurements couple the $n$ resonances to the incoming and outgoing planewaves, whose complex amplitudes we denote by two $2 \times 1$ column vectors, $s_+$ and $s_-$. The direct reflection and transmission of the planewaves through the slab (in the absence of resonances) are described by a $2 \times 2$ complex symmetric matrix $C$, and
\begin{equation}
\label{eq:CMT_2}
s_- = C s_+ + D A,
\end{equation}
where $D$ and $K$ in equation~\eqref{eq:CMT_1} are $2 \times n$ complex matrices denoting coupling between the resonances and the planewaves. 
We approximate the direct scattering matrix $C$ by that of a homogeneous slab whose permittivity is equal to the spatial average of the PhC slab\cite{2002_Fan_PRB, 2003_Fan_JOSAA, 2013_Hsu_Nature}.
Lastly, outgoing planewaves into the silica substrate are reflected at the silica-silicon interface, so
\begin{equation}
\label{eq:CMT_3}
s_{2+} = e^{2 i \beta h_s} r_{23} s_{2-},
\end{equation}
where $h_s$ is the thickness of the silica substrate with refractive index $n_{\rm s} = 1.46$, $\beta = \sqrt{n_{\rm s}^2 \omega^2/c^2 - |{\bf k}_{\parallel}|^2}$ is the propagation constant in silica, and $r_{23}$ is the Fresnel reflection coefficient between silica and the underlying silicon.
The formalism described above is the same as Ref.~\citen{2013_Hsu_Nature} except that here we describe the $n$ resonances in a more general setting that accounts for their coupling (off-diagonal terms of $H$) and therefore their non-orthogonality.


For steady state with $e^{-i \omega t}$ time dependence, we solve for vector $A$ from equation~\eqref{eq:CMT_1} to get the scattering matrix of the whole system that includes both direct and resonant processes,
\begin{equation}
\label{eq:ss}
s_- = (C + C_{\rm res}) s_+, 
\end{equation}
where the effect of the $n$ resonances is captured in a $2 \times 2$ matrix
\begin{equation}
\label{eq:Cres_1}
C_{\rm res} = i D ( \omega - H )^{-1} K^{\rm T}.
\end{equation}
We can solve equation~\eqref{eq:CMT_3} and equation~\eqref{eq:ss} to obtain the reflectivity
\begin{equation}
\label{eq:R}
R_{\rm TCMT} = \left| \frac{s_{1-}}{s_{1+}} \right|^2. 
\end{equation}
In this expression, the only unknown is $C_{\rm res}$. Therefore, by comparing the experimentally measured reflectivity spectrum $R(\omega)$ and the one given by TCMT in equation~\eqref{eq:R}, we can extract the unknown parameters in the resonant scattering matrix $C_{\rm res}$ and obtain the eigenvalues of the Hamiltonian $H$.

The remaining task is to write $C_{\rm res}$ using as few unknowns as possible so that the eigenvalues of $H$ can be extracted unambiguously. In equation~\eqref{eq:Cres_1}, there are a large number of unknowns in the matrix elements of $H$, $D$, and $K$, but there is much redundancy because the matrix elements are not independent variables and because $C_{\rm res}$ is independent of the basis choice.
Below, we show that we can express $C_{\rm res}$ with only $2n + 1$ unknown real numbers, and these $2n + 1$ real numbers are enough to determine the $n$ complex eigenvalues of $H$.

First, we normalize the amplitudes of $A$ and $s_{\pm}$ such that their magnitude squared are the energy of the resonances per unit cell and the power of the incoming/outgoing planewaves per unit cell, respectively.
Then, energy conservation, time-reversal symmetry, and $C_2$ rotational symmetry of the PhC slab\cite{2004_Suh_IEEE, 2013_Hsu_LSA} require the direct scattering matrix to satisfy $C^\dagger = C^* = C^{-1}$ and the coupling matrices to satisfy $D^\dagger D = 2 \Gamma \label{eq:DDG}$,  $K = D$, and  $C D^* = -D$.
It follows that the matrix $\Gamma$ is real and symmetric.
Next, using the Woodbury matrix identity and these constrains, we can rewrite equation~\eqref{eq:Cres_1} as
\begin{equation}
\label{eq:Cres_2}
C_{\rm res} =  - 2 W \left( 2 + W\right)^{-1} C,
\end{equation}
where $W \equiv i D (\omega - \Omega + i \gamma_0)^{-1} D^\dagger$ is a 2-by-2 matrix. We note that the matrix $W$, and therefore the matrix $C_{\rm res}$, is invariant under a change of basis for the resonances through any orthogonal matrix $U$ (where $\Omega$ is transformed to $U\Omega U^{-1}$, and $D$ is transformed to $D U^{-1}$). Therefore, we are free to choose any basis. Given the expression for $W$, we choose the basis where $\Omega$ is diagonal, so $\Omega_{ij} = \Omega_j \delta_{ij}$, with $\{\Omega_j\}_{j=1}^n$ being the eigenvalues of $\Omega$. 

To proceed further, we note that the PhC slab sits on a silica substrate with $n_{\rm s} = 1.46$ and is immersed in a liquid with $n=1.48$, so the structure is nearly symmetric in $z$ direction. The mirror symmetry requires the coupling to the two sides to be symmetric or anti-symmetric\cite{JJ_book},
\begin{equation}
\label{eq:d_sym}
\frac{D_{1j}}{D_{2j}} \equiv \sigma_j = \pm 1, \quad j=1, \ldots, n,
\end{equation} 
where $\sigma_j = 1$ for TE-like resonances and $\sigma_j = -1$ for TM-like resonances, in the convention where $(E_x, E_y)$ determines the phase of $A_j$ and $s_{\pm}$.
Then, the diagonal elements of $\Gamma$ are related to $D$ by $\Gamma_{jj} \equiv \gamma_j = |D_{1j}|^2$, and in this basis we have
\begin{equation}
\label{eq:DWD_2}
W =  \sum_{j=1}^{n} \frac{i \gamma_j}{\omega - \Omega_j + i \gamma_{\rm nr}}
\begin{pmatrix}
1 & \sigma_j \\
\sigma_j & 1
\end{pmatrix}.
\end{equation}
This completes our derivation. Equations~\eqref{eq:Cres_2} and \eqref{eq:DWD_2} provide an expression for $C_{\rm res}$ that depends only on $2n+1$ unknown non-negative real numbers: the $n$ eigenvalues $\{\Omega_j\}_{j=1}^n$ of the Hermitian matrix $\Omega$, the $n$ diagonal elements $\{\gamma_j\}_{j=1}^n$ of the real-symmetric radiation matrix $\Gamma$ in the basis where $\Omega$ is diagonal, and the non-radiative decay rate $\gamma_{\rm nr}$.

At each angle and each polarization, we fit the experimentally measured reflectivity spectrum $R(\omega)$ to the TCMT expression equation~\eqref{eq:R} to determine these $2n+1$ unknown parameters. Fig.~3a and Fig.~S6a show the comparison between the experimental reflectivity spectrum and the fitted TCMT reflectivity spectrum at some representative angles. The near-perfect agreement between the two demonstrates the validity of the TCMT model.

To obtain the eigenvalues of the Hamiltonian $H$, we also need to know the off-diagonal elements of $\Gamma$.
From $D^\dagger D = 2\Gamma$ and $D_{1j}/D_{2j} = \sigma_j$, we see that $\Gamma_{ij} = 0$ when resonance $i$ and resonance $j$ have different symmetries in $z$ ({\it i.e.} when $\sigma_i \sigma_j \neq 1$), and that $\Gamma_{ij} = \pm \sqrt{\gamma_{i} \gamma_{j}}$ when $\sigma_i \sigma_j = 1$. In the latter case, the sign of $\Gamma_{ij}$ depends on the choice of basis; the eigenvalues of $H$ are independent of the basis choice, so to calculate the eigenvalues of $H$, we can simply take the positive root for all of the non-zero off-diagonal elements of $\Gamma$.

We note that the model Hamiltonians introduced previously, such as equation~\eqref{eqn:NH-Hamiltonian} in the main text and equations \eqref{eq:Heff3dU} and \eqref{eqn:neardegeneracy-with-loss} above, are all special cases of the general Hamiltonian in equation~\eqref{eq:H_TCMT} that we consider in the TCMT formalism in this section.
Those model Hamiltonians fix the number of resonances, assume simple forms of their parameters, and choose a specific basis in order to convey the physical picture.
Meanwhile, the TCMT formalism in this section does not make such assumptions (aside from basic principles such as energy conservation and time-reversal symmetry) so that it can be used as an unbiased method for analyzing the experimental data.

We also note that the TCMT equations, from \eqref{eq:CMT_1} to \eqref{eq:d_sym}, are all written in the general matrix notation where one is free to choose any basis for the Hamiltonian $H$; we only make the specific basis choice (the basis where $\Omega$ is diagonal) in equation~\eqref{eq:DWD_2} in order to simplify the expression for matrix $W$,
and in equation~\eqref{eq:TCMT} of the main text in order to emphasize the eigenvalues $\Omega$.
Meanwhile, physical observables, such as $C_{\rm res}$ in equation~\eqref{eq:Cres_1}, $R_{\rm TCMT}$ in equation~\eqref{eq:R}, and the eigenvalues of $H$, are all independent of the basis choice.

\subsection{Section VI, Reflection peaks and CRIT\\} 


In this section, we use a simplified scenario (a special case of the previous section) to illustrate that the peaks of the reflectivity generally follow the eigenvalues of $\Omega$ and to show the coupled-resonator-induced-transparency (CRIT).

Consider a simplified scenario with two resonances of the same symmetry in $z$ and without non-radiative loss ({\it i.e.} $n=2$, $\sigma_1 = \sigma_2$, $\gamma_{\rm nr} = 0$), and ignore the direct Fresnel reflection between the dielectric layers (so that the direct scattering matrix $C$ has no reflection, and that $s_{2+} = 0$ in equation~\eqref{eq:CMT_3}). In such case, equations~\eqref{eq:R} \eqref{eq:Cres_2} \eqref{eq:DWD_2} give
\begin{equation}
\label{eqn:reflectivity}
R_{\rm TCMT}(\omega) = \frac{1}{1+f^2(\omega)}, \quad \frac{1}{f(\omega)} = \frac{\gamma_1}{\omega-\Omega_1} + \frac{\gamma_2}{\omega-\Omega_2}.
\end{equation}
We immediately see that the reflectivity reaches its maximal value of 1 when $\omega = \Omega_1$ or $\omega = \Omega_2$, namely at the eigenvalues of the matrix $\Omega$.
Another feature we can observe is that the reflectivity is 0 when $\omega = ({\gamma_{1}\Omega_{2}+\gamma_{2}\Omega_{1}})/({\gamma_{1}+\gamma_{2}})$, which is a phenomenon called coupled-resonator-induced-transparency (CRIT)\cite{2010_Miroshnichenko_RMP, 2014_Hsu_NL}.

We emphasize that the reflectivity peaks are different from the complex eigenvalues of the Hamiltonian $H$.
Consider a simple example with $\Omega_{1,2} = \omega_{0} \pm b$, and $\gamma_{1}=\gamma_{2}=b$. The reflection peaks at $\Omega_{1,2}=\omega_{0} \pm b$, while the two complex eigenvalues are degenerate at $\omega_{+} = \omega_{-} = \omega_{0} - i b$, whose real part is in the middle of the two reflection peaks. This explains the reflectivity from the PhC slabs at $0.3^{\circ}$ shown in Fig.~2d and Fig.~3a, where the degenerate complex eigenvalues of the system are in between the two reflection peaks. 


In Fig.~S5, we use some examples to illustrate the difference between the reflectivity peaks and the complex eigenvalues.
Fig.~S5a shows the case when there is only one resonance (removing one of the two terms in equation~\eqref{eqn:reflectivity}) with complex eigenvalue $\omega_{0} - i\gamma$; in this case, the reflection peak (red arrow) is at the same position as the real part of the complex eigenvalue.
In contrast, Fig.~S5b shows the case when there are two resonances (equation~\eqref{eqn:reflectivity}) with $\Omega_{1,2}$ fixed at $\omega_{0} \pm b$; as we vary $\gamma_{1,2}$, the complex eigenvalues (circles) vary accordingly, whereas the reflectivity peaks (red arrows) always show up at $\Omega_{1,2}$. 

For the realistic PhC slab structure in our experiment, the reflectivity is described by the more general expression, equations~\eqref{eq:R}, but equation~\eqref{eqn:reflectivity} serves as a qualitative approximation near the frequency range of interest, because the far-away resonances do not contribute much, the non-radiative loss is small, and the Fresnel reflection between the dielectric layers (liquid, Si$_3$N$_4$, silica, and silicon) is small. So, we can still see the general trend that the reflectivity peaks follow the eigenvalues of $\Omega$ (as evident by comparing Fig.~S3 and Fig.~S6b), and we can still see reflectivity dips for CRIT (such as in Fig.~3a).

\clearpage

\renewcommand{\figurename}{{FIG.}}
\renewcommand{\thefigure}{{ S\arabic{figure}}}
\setcounter{figure}{0}

\begin{figure*}[ht]
\centerline{
\includegraphics[width=0.8\textwidth]{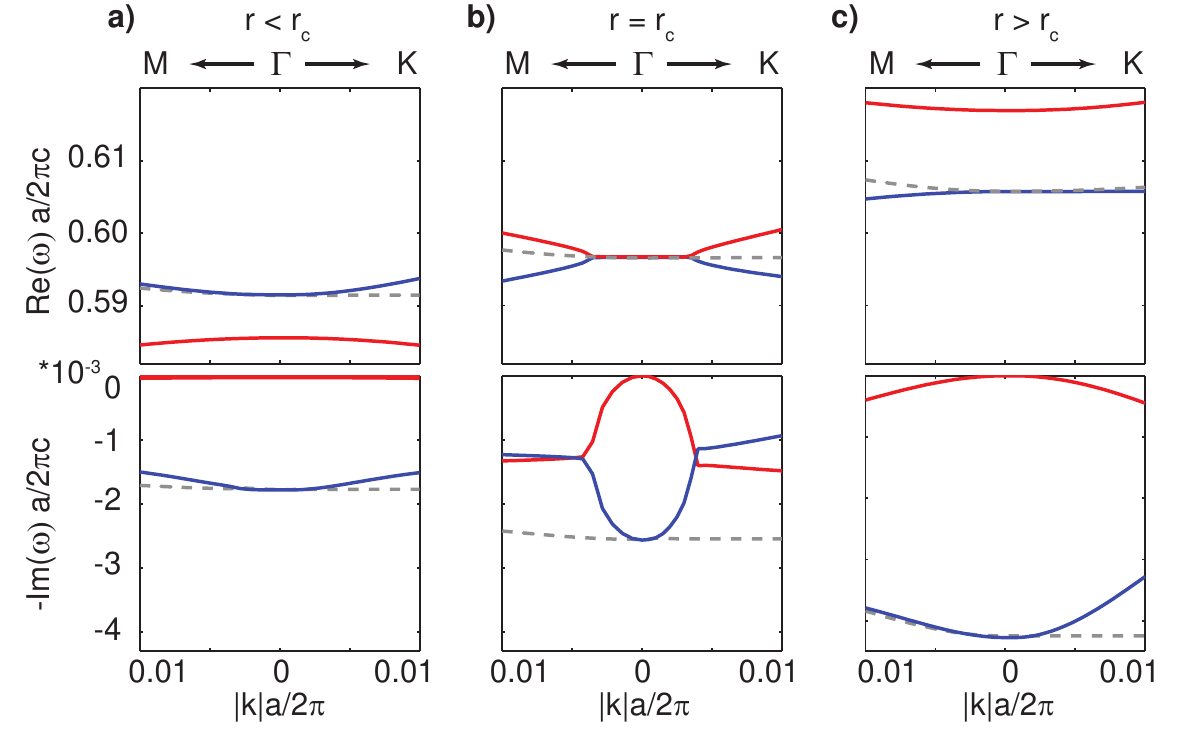}
}\caption{\baselineskip20pt{\bf Simulation results of the complex eigenvalues of the PhC slabs with and without accidental degeneracy.}
The real (upper panels) and imaginary (lower panels) parts of the complex eigenvalues are shown for structures with ({\bf b}) and without ({\bf a, c}) accidental degeneracy. The bands with quadrapole modes in the middle of the Brillouin zone are shown in red solid lines, while the bands with the dipole mode are shown in blue solid lines and gray dashed lines. The bands shown in red and blue solid lines couple to each other, while the band in gray dashed lines is decoupled from the other two due to symmetry. When accidental degeneracy happens (at $r=r_{\rm c}$ as shown in {\bf b}), the characteristic branching features are observed demonstrating the existence of EPs. When the accidental degeneracy is lifted, the quadrupole band splits from the dipole bands from different directions: from the bottom when radius of the holes is too small  ($r<r_{\rm c}$ as shown in {\bf a}), or from the top when the radius is too big ($r>r_{\rm c}$ as shown in {\bf c}).
}
\label{figS1}
\end{figure*}

\begin{figure*}[ht]
\includegraphics[width=\textwidth]{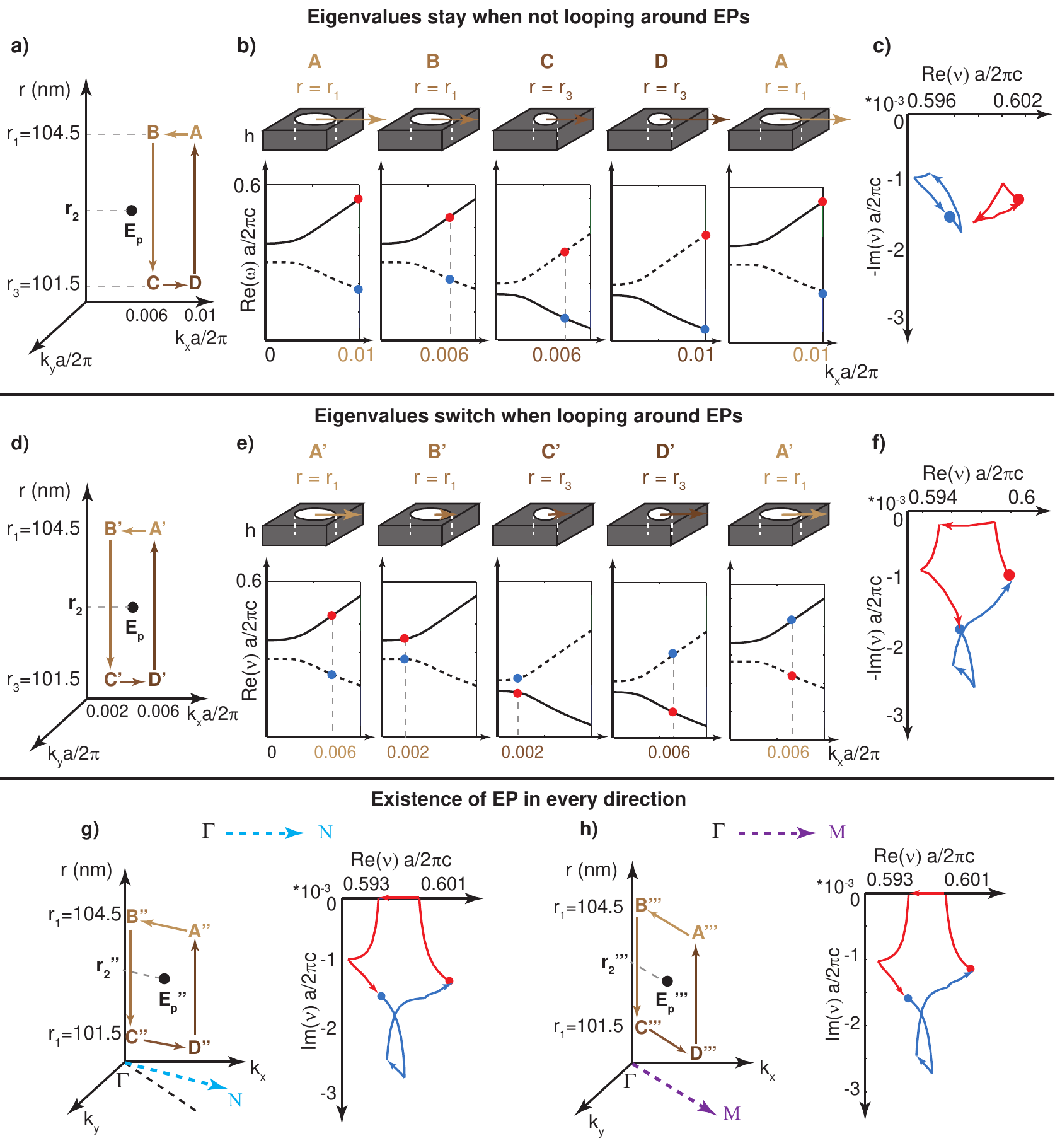}
\label{figS2}
\end{figure*}

\clearpage
{\baselineskip16pt {FIG. S2. {\bf  Existence of EP along every direction in the momentum space for the realistic PhC slab structure.}}
{\bf a}, A loop is created in the parameter space of the structure $(A\rightarrow B\rightarrow C\rightarrow D\rightarrow A)$, which does not enclose the EP of the system (point ${\rm E_{p}}$). Here, $r$ is the radius of the air holes, and $k_{x,y}$ are the in-plane wavevectors. 
{\bf b}, For each point along the loop, we numerically calculate the eigenvalues of the PhC slab with the corresponding hole radius at the corresponding in-plane wavevector. 
{\bf c}, 
The complex eigenvalues return to their initial positions at the end of the loop (namely, the blue dot and the red dot come back to themselves) when the system parameters come back to point A. 
{\bf d,e,f}, Another loop is created $(A'\rightarrow B'\rightarrow C'\rightarrow D'\rightarrow A')$, which encloses an EP of the system (the same point ${\rm E_{p}}$ as in a.). 
Following this new loop, the two eigenvalues switch their positions at the end of the loop (namely, the blue dot and the red dot switch their positions) when the system parameters come back to point $A'$. 
{\bf g,h}, The two complex eigenvalues always switch their positions when we choose the right loops along other directions in the momentum space ($\Gamma$ to N in {\bf g} and $\Gamma$ to M in {\bf h}). 
}

\setcounter{figure}{2}

\begin{figure*}[ht]
\centerline{
\includegraphics[width=0.9\textwidth]{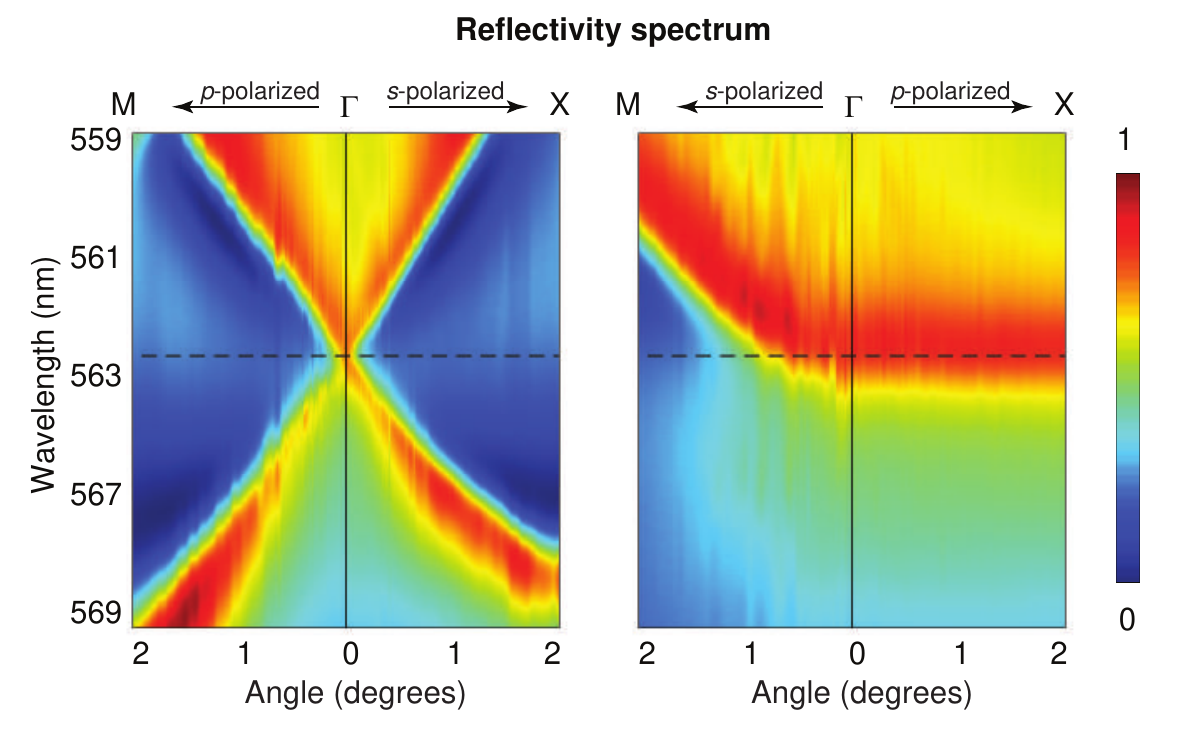}
}\caption{\baselineskip20pt{\bf Experimental results of reflectivity showing an accidental Dirac cone.}
Light with different polarizations  ($s$ and $p$) is selected to excite different resonances of the PhC slab along different directions in the $k$ space. Depending on the choice of polarization, the two bands forming the conical dispersion are excited ($s$-polarized along $\Gamma$-X and $p$-polarized along $\Gamma$-M, shown in the left panel), or the flat band in the middle is excited ($p$-polarized along $\Gamma$-X and $s$-polarized along $\Gamma$-M, shown in the right panel).
}
\label{figS6}
\end{figure*}

\begin{figure*}[ht]
\includegraphics[width=\textwidth]{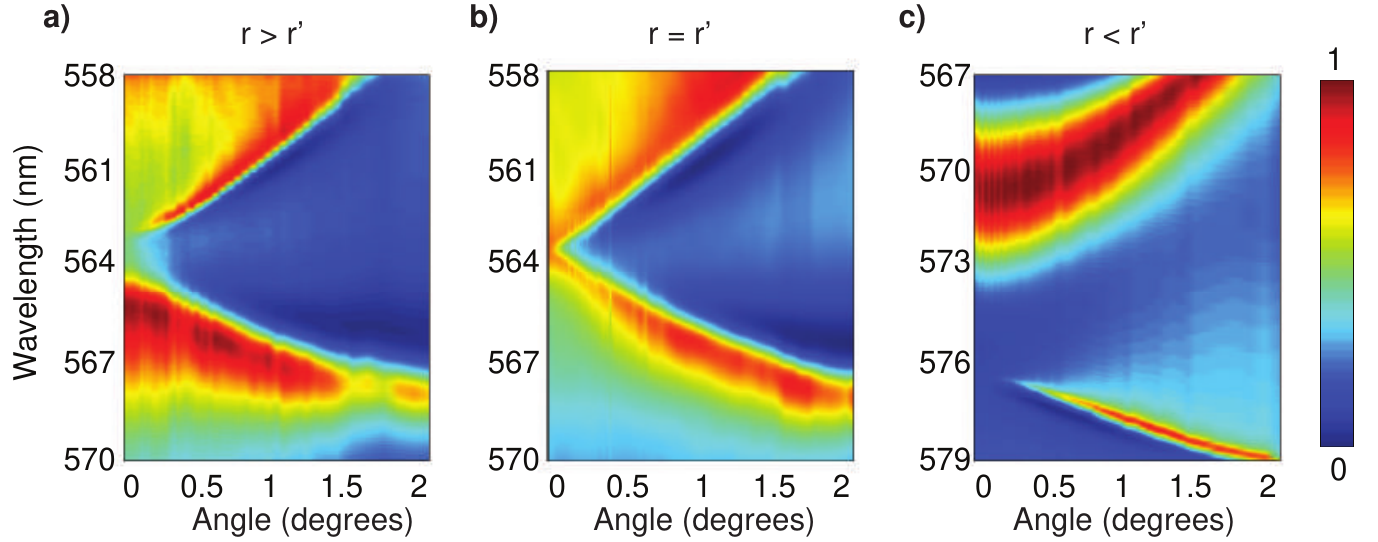}
\caption{\baselineskip20pt{\bf Experimental results of reflectivity from PhC slabs with and without accidental degeneracy.}
Angle-resolved reflectivity along the $\Gamma$ to X direction, measured for three different PhC slabs: {\bf a}, with smaller hole radius than the structure with accidental degeneracy ($r<r'$); {\bf b}, with accidental degeneracy ($r=r'$); {\bf c}, with bigger hole radius than the structure with accidental degeneracy ($r>r'$). The reflectivity peaks of the structures without accidental degeneracy (a,c) follow quadratic dispersions; while the reflectivity peaks of the structure with accidental degeneracy (b) follow linear Dirac dispersion. Data shown in (b) is the same data as in Fig.~2c the left panel of Fig.~S3. }
\label{figS3}
\end{figure*}

\begin{figure*}[ht]
\centerline{
\includegraphics[width=0.4\textwidth]{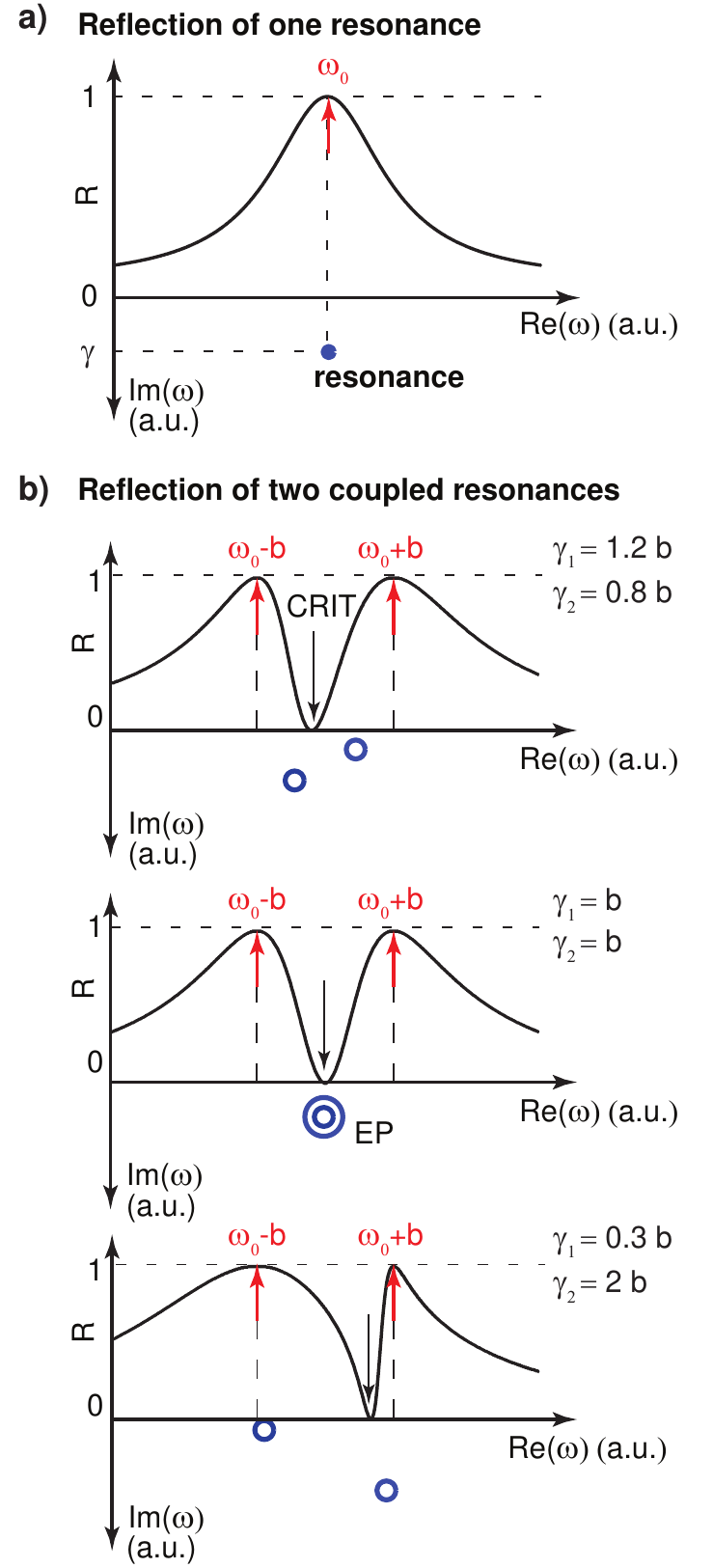}
}
\caption{\baselineskip20pt{\bf Illustrative reflectivity spectrum from one resonance and from two coupled resonances.}
{\bf a}, When a single resonance dominates, the reflectivity peak is at the same position as the real part of the complex eigenvalue. 
{\bf b}, With two coupled resonances, the reflectivity peaks (red arrows) no longer follow the eigenvalues of the system (blue circles).
As we vary the radiation loss $\gamma_{1,2}$ of the two resonances while fixing the eigenvalues $\Omega_{1,2}$ of the Hermitian matrix, we see the complex eigenvalues vary whereas the reflectivity peaks are fixed.
The middle panel of (b) shows a situation when the two complex eigenvalues coalesce into an EP. 
}
\label{figS4}
\end{figure*}

\begin{figure*}[ht]
\includegraphics[width=\textwidth]{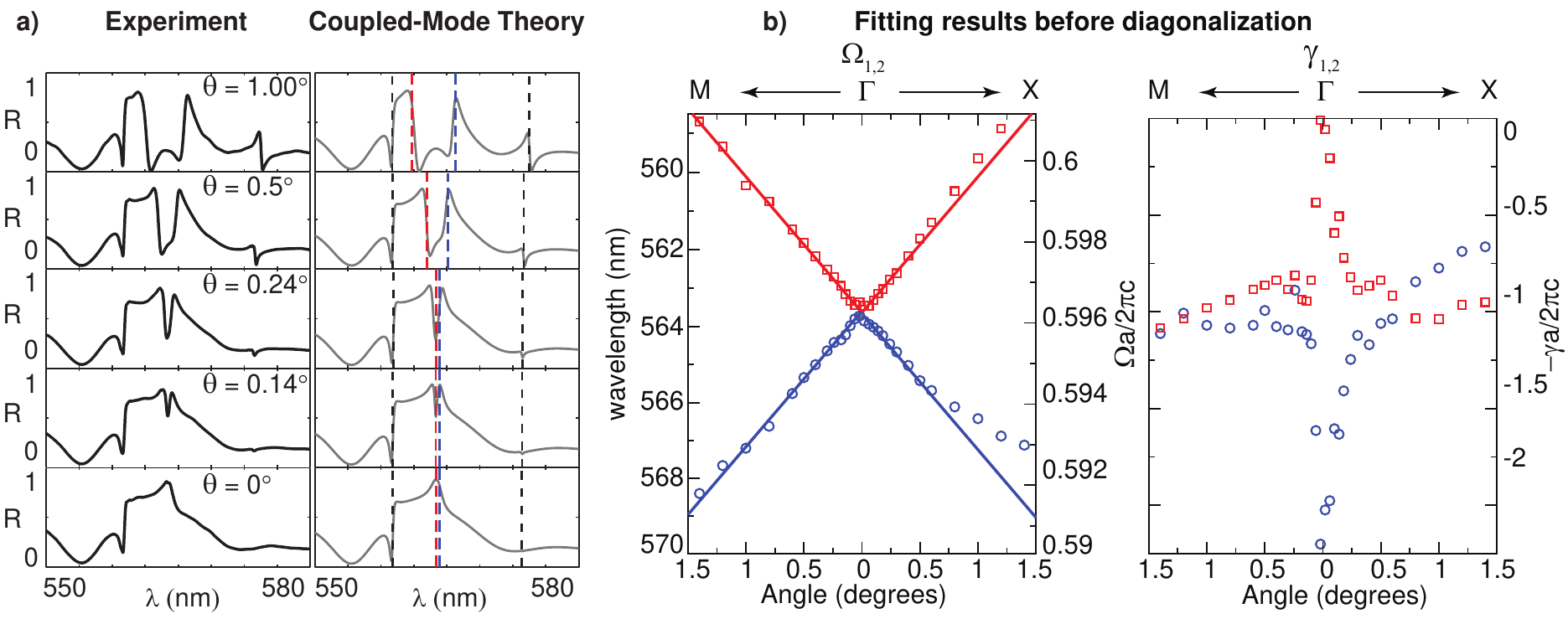}
\caption{\baselineskip20pt{\bf TCMT fitting and visualization of accidental Dirac cone.}
{\bf a}, Examples of reflection spectrum measured at five different incident angles ($0^{\circ}$, $0.14^{\circ}$, $0.24^{\circ}$, $0.5^{\circ}$ and $1^{\circ}$) along the $\Gamma$-X direction for $s$ polarization, with comparison to the TCMT expression in equation~\eqref{eq:R} after fitting. Dotted lines indicate the resonances, with the two relevant resonances marked in red and blue.
{\bf b}, Parameters obtained from the TCMT fitting. The eigenvalues for the Hermitian part of the Hamiltonian, $\Omega_{1,2}$, are shown in the left panel and reveal the Dirac dispersion arising from accidental degeneracy. The diagonal terms for the anti-Hermitian part of the Hamiltonian, $\gamma_{1,2}$, are shown in the right panel. Note that the anti-Hermitian part of the Hamiltonian also has off-diagonal terms, so $\Omega_{1,2}$ and $\gamma_{1,2}$ are not the eigenvalues of the Hamiltonian. The eigenvalues of the Hamiltonian are shown in Fig.~3 of the main text.
}
\label{figS5}
\end{figure*}


\end{document}